\documentclass{IEEEtran}
\usepackage{graphicx,
	psfrag,
	epsfig,
	epstopdf,
	amsthm,
    amsmath,
	amssymb,
	url,
	subfigure,
	algorithm,
	algorithmic,
	balance,
	enumerate,
	color,
    url,
	setspace,
	tikz,
    placeins,
}
\usepackage{cite}
\usepackage{microtype}

\newcommand{\cref}[1]{Constraint~\ref{#1}}

\newcommand{\ignore}[1]{}

\makeatletter
 \def\@textbottom{\vskip \z@ \@plus 4pt}
 \let\@texttop\relax
\makeatother

\begin{document}
\title{Edge-enabled V2X Service Placement for Intelligent Transportation Systems}	
\author{
\IEEEauthorblockN{Abdallah Moubayed\IEEEauthorrefmark{1}, Abdallah Shami\IEEEauthorrefmark{1}, Parisa Heidari\IEEEauthorrefmark{2}, Adel Larabi \IEEEauthorrefmark{2}, and Richard Brunner \IEEEauthorrefmark{2}} \\
\IEEEauthorblockA{\IEEEauthorrefmark{1} Western University, London, Ontario, Canada; e-mails: \{amoubaye, abdallah.shami\}@uwo.ca\\
}
\IEEEauthorblockA{\IEEEauthorrefmark{2}  Edge Gravity by Ericsson, Montreal, Quebec, Canada; e-mails:\{PHeidari, ALarabi, RBrunner\}@edgegravity.ericsson.com}	
}
\maketitle
\begin{abstract}
Vehicle-to-everything (V2X) communication and services have been garnering significant interest from different stakeholders as part of future intelligent transportation systems (ITSs). This is due to the many benefits they offer. However, many of these services have stringent performance requirements, particularly in terms of the delay/latency. Multi-access/mobile edge computing (MEC) has been proposed as a potential solution for such services by bringing them closer to vehicles. Yet, this introduces a new set of challenges such as where to place these V2X services, especially given the limit computation resources available at edge nodes. To that end, this work formulates the problem of optimal V2X service placement (OVSP) in a hybrid core/edge environment as a binary integer linear programming problem. To the best of our knowledge, no previous work considered the V2X service placement problem while taking into consideration the computational resource availability at the nodes. Moreover, a low-complexity greedy-based heuristic algorithm named ``Greedy V2X Service Placement Algorithm'' (G-VSPA) was developed to solve this problem. Simulation results show that the OVSP model successfully guarantees and maintains the QoS requirements of all the different V2X services. Additionally, it is observed that the proposed G-VSPA algorithm achieves close to optimal performance while having lower complexity. 
\end{abstract}

\section{Introduction}\label{intro}
\indent As part of the development and deployment efforts of intelligent transportation systems (ITSs), different stakeholders ranging from governmental agencies to automotive manufacturers have been significantly interested in vehicle-to-everything (V2X) communication  \cite{v2x_interest1,v2x_interest2}. This is because the many projected benefits of such systems including reduction in traffic-related accidents, introduction of new business models, and a decrease in operational expenditures of vehicular fleets \cite{v2x_access_survey1,v2x_access_survey2}. Moreover, V2X communications is required to offer a variety of services such as autonomous vehicle operation, traffic flow optimization, and in-car entertainment \cite{v2x_communication_modes}. Some of these services and applications have stringent performance requirements, particularly those related to end-to-end (E2E) latency/delay. For example, pre-sense crash warning has a minimum requirement of 20-50 ms E2E latency as per the US department of transportation and the European Telecommunications Standards Institute (ETSI) \cite{traffic_safety1,traffic_safety2,traffic_safety3}.\\
\indent To address the challenge of guaranteeing E2E latency, multi-access/mobile edge computing (MEC), one component of fog computing paradigm, has been proposed as a potential solution. In laymen terms, edge computing refers to making computing resources and storage available in a nearby proximity to sensors and mobile devices. In the context of ITSs and V2X communications, MEC has been introduced as an extension to cloud computing paradigm to minimize the latency of serving data through hosting the applications and services on servers with the closest proximity to the end-users. This in turn can help providers support real-time applications, mobility, and location-aware services.\\
\indent However, adopting a distributed cloud/edge computing paradigm brings a new set of challenges. One such challenge is where to place the V2X applications and services. This is because despite the lower latencies offered by hosting such applications and services on edge computing nodes, such nodes tend to have limited computation and storage capabilities. Another challenge is how to ensure the high availability of these applications and services. This is of particular interest, especially for autonomous driving and traffic safety applications to avoid accidents and optimize traffic flow. Therefore, this work focuses on addressing these two challenges by formulating an optimization problem that aims to find a V2X application/service placement that minimizes the E2E latency. The problem takes into consideration delay, computational, and high availability requirements of such applications/services given the available computational resources when making the placement decision.\\
\indent A summary of this work's contributions are:
\begin{itemize}
	\item Formulate the problem of optimal V2X service placement (OVSP) in a hybrid cloud/edge environment as a binary integer linear programming problem.
	\item Develop a low-complexity greedy-based heuristic algorithm titled ``Greedy V2X Service Placement Algorithm'' (G-VSPA).
	\item Evaluate the performance of the proposed OVSP model and G-VSPA algorithm in terms of average delay/latency and average node computing utilization.  
\end{itemize}
To the best of our knowledge, no previous work considered the V2X service placement problem while taking into consideration the availability of computational resources at the core and edge nodes of a hybrid computing environment.
\begin{figure*}[!tb]
	\centering
	\includegraphics[scale=.6]{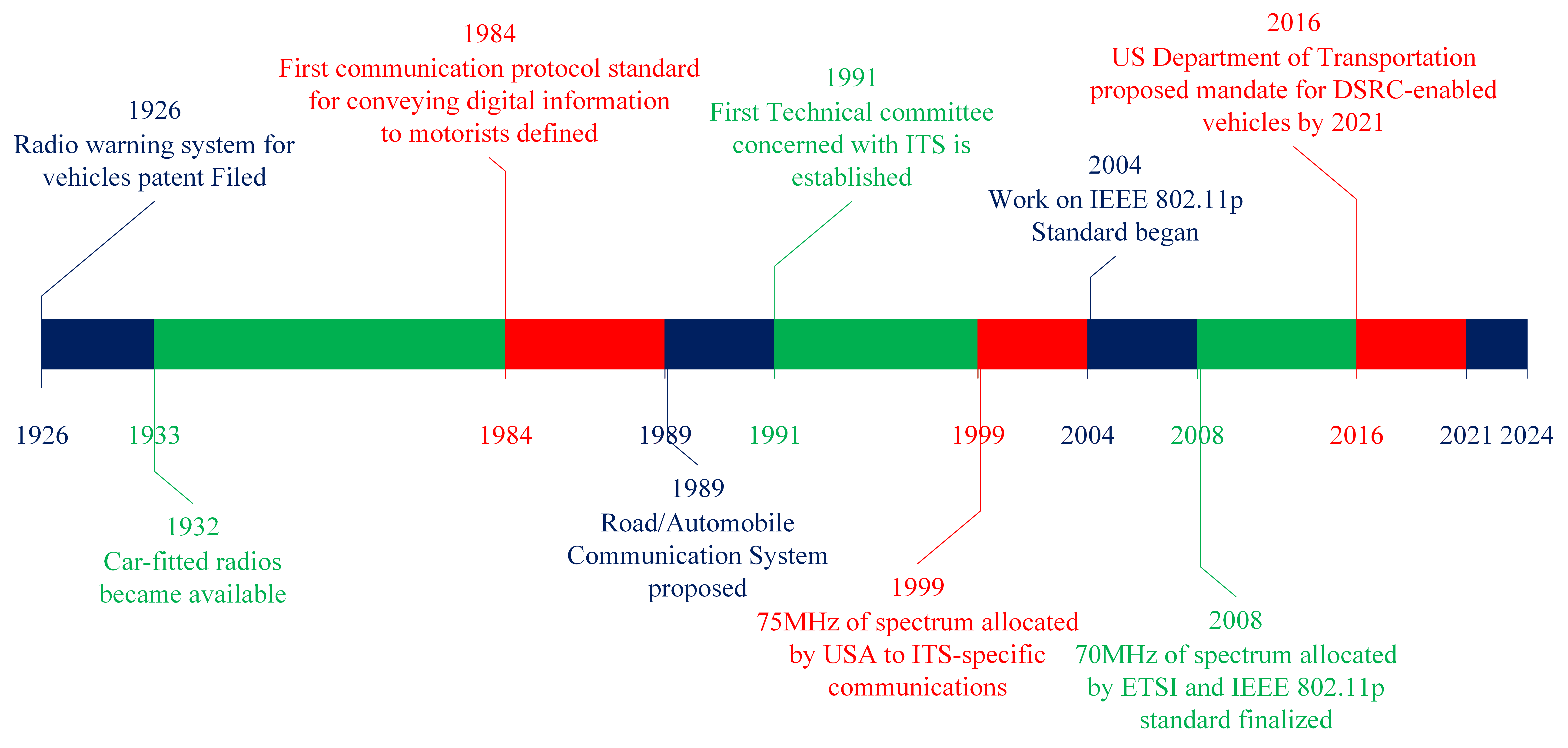}
	\caption{V2X Communication Timeline}
	\label{v2x_history}
\end{figure*}

\indent The remainder of this paper is organized as follows: Section \ref{background} provides a brief background about both V2X communication (history, communication modes, and applications) and multi-access/mobile edge computing. Section \ref{related_work} provides a brief discussion of some of the previous work conducted within this area. Section \ref{sys_model} presents the system model considered in this work. Section \ref{prob_formulation} formulates the optimization problem. Section \ref{gvspa} describes the proposed greedy-based G-VSPA heuristic algorithm and discusses its complexity. Section \ref{performance_eval} provides a performance evaluation and discussion of the system. Finally, Section \ref{conc} concludes the paper.

\section{Background}\label{background}
\subsection{V2X Communication:}
\indent V2X communication is a core component of modern ITS systems. V2X communication governs the communication and coordination between vehicles and their environment. This includes communication between vehicles and other entities typically found on the road such as other vehicles, pedestrians, and infrastructure. This results in having a more economical, efficient, and safe autonomous overland transportation system. Figure \ref{v2x_history} provides a brief summary of the key milestones in the development of V2X technologies and applications. In what follows, an overview of the communication modes adopted in V2X systems and their applications is presented. 
\begin{figure*}[!tb]
	\centering
	\includegraphics[scale=0.6]{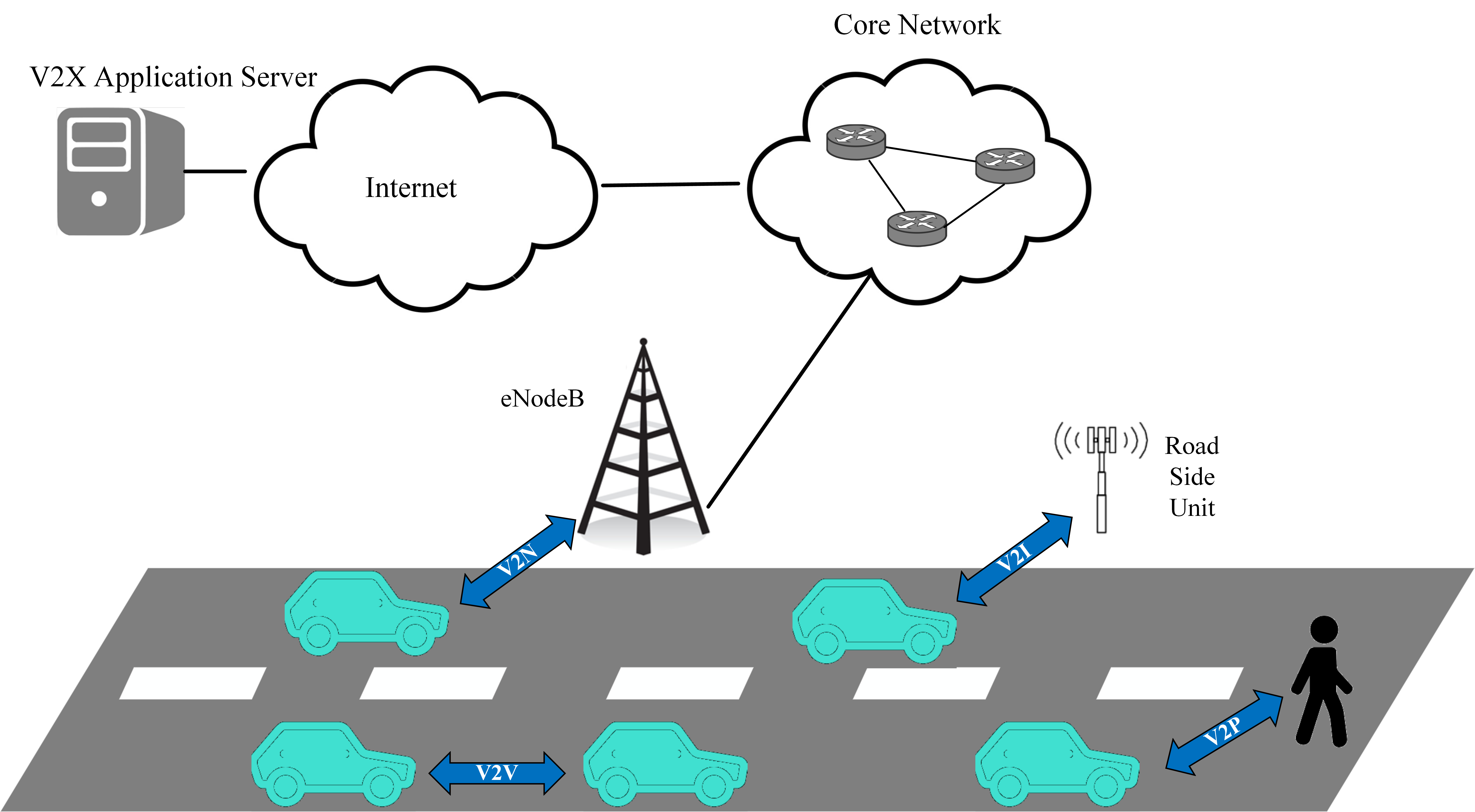}
	\caption{V2X Communication Modes}
	\label{comm_modes}
\end{figure*} 
\subsubsection{Communication Modes}
\indent To cover all possible on-road interactions, the 3GPP project proposed four different communication modes. This includes vehicle-to-network (V2N), vehicle-to-infrastructure (V2I), vehicle-to-vehicle (V2V), and vehicle-to-pedestrian (V2P) communication \cite{v2x_communication_modes} as shown in Fig. \ref{comm_modes}. Depending on the service or application, a communication mode can be chosen. A brief overview of each of these communication modes is provided below:
\begin{enumerate}[a-]
	\item V2N Communication: V2N communication refers to the communication between a vehicle and a V2X application server. This is typically done using a cellular network such as an LTE network \cite{C-V2X1,C-V2X2}. Through this connection, different services such as infotainment, traffic optimization, navigation, and safety can be offered \cite{v2n_applications1,v2n_applications2}. 
	\item V2I Communication: V2I communication refers to the communication between a vehicle and roadside infrastructure such as road-side units (RSUs). This mode  is typically used to disseminate safety messages to multiple drivers within the RSU's coverage area. Additionally, it can be used to share information at signalized and non-signalized intersections to avoid collisions \cite{v2i_applications1,v2i_applications2}.
	\item V2V Communication: V2V communication refers to the direct communication between two vehicles. Cooperative driving is enabled with this communication mode by exchanging  different messages such as collision warning/avoidance, lane change warning, and emergency vehicle warning \cite{v2v_applications1,v2v_applications2}.
	\item V2P Communication: V2P communication refers to the direct communication between vehicles and vulnerable road users (VRUs) such as pedestrians, wheelchair users, bikers, and motorcyclists. Again, this communication mode can be used for road safety. More specifically, both the vulnerable users and the vehicle and the VRU are alerted of a possible collisions using this mode \cite{v2p_applications1,v2p_applications2}.
\end{enumerate}
\subsubsection{Applications}\label{V2X_applications_list}
\indent V2X communications enables a variety of applications and services. Each of these applications and services have different throughput, latency, and frequency requirements. Accordingly, these applications and services are often grouped into four main categories:
\begin{enumerate}[a-]
	\item Autonomous/Cooperative Driving: The first category is autonomous and cooperative driving which mainly focuses on V2V communication between vehicles in close proximity. This particular application has extremely stringent requirements in terms of the communication throughput and latency. More specifically, such an application requires a throughput $\geq$ 5 Mbps and latency $\leq$ 10 ms \cite{autonomous_driving1,autonomous_driving2}.
	\item Traffic Safety: The second category of applications is overall traffic safety. This represents a more general view of the autonomous/cooperative driving application. Traffic safety applications have many objectives including: reduction in the number and severity inter-vehicle collisions, protection of vulnerable road users, and reduction of property damage. As can be expected, such applications have very strict requirements. For example, the pre-sense crash warning has a minimum requirement of 20-50 ms of round-trip latency \cite{traffic_safety1,traffic_safety2}. Additionally, the throughput required for some traffic safety services such as road sign recognition is estimated to be 700 Mbps \cite{traffic_safety3}.
	\item Traffic Efficiency: The third category of V2X applications and services is traffic efficiency. This category of applications focus on various tasks such as coordinating intersection timings, planning the route from source to destination for various vehicles, and sharing general information including geographical location and road conditions. Such applications often have less strict requirements. For example, the tolerable latencies for such applications ranges between 100-500 ms and the throughput ranges between 10-45 Mbps \cite{traffic_safety2}.
	\item Infotainment: The fourth category is infotainment applications. This refers to the set of services that aim to provide general non-driving related information (location of car rental services) and entertainment (video streaming). Such applications and services typically have the lowest requirement. For example, latencies up to 1 sec can be tolerated. Furthermore, the throughput requirement is estimated to be around 80 Mbps, which is comparable to that of conventional mobile services \cite{autonomous_driving1,traffic_safety2}.
\end{enumerate}
\begin{figure*}[!tb]
	\centering
	\includegraphics[scale=0.6]{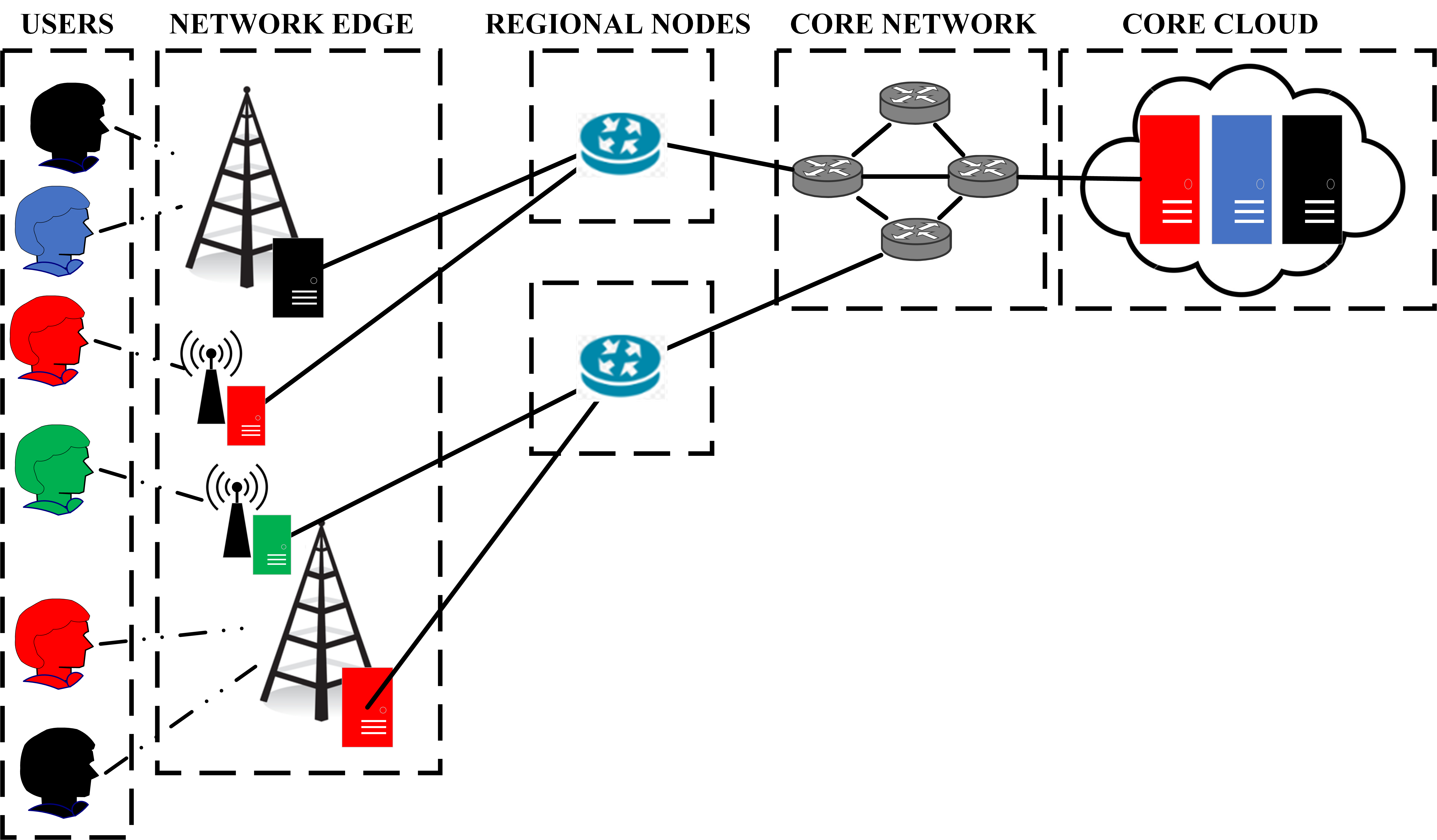}
	\caption{General MEC Architecture}
	\label{MEC_arch}
\end{figure*}
\subsection{Multi-access/Mobile Edge Computing:}\label{MEC_background}
\indent Multi-access/Mobile Edge computing (MEC) has been discussed as a potential solution and key enabler of future networks and systems. By bringing more computational power closer to the users, the goal of MEC is to reduce the experienced latency as well as decrease the signaling overhead in the cloud core. In what follows, a brief description of the MEC concept and its motivation.
\subsubsection{Concept \& Motivation}\mbox{}\\
\indent The MEC concept was first introduced as an extension to the cloud computing paradigm . The idea is to shift some of the computational resources to the edge of the network \cite{ETSI_MEC}, as shown in Fig. \ref{MEC_arch}. This was done in response to the continued growth of connected devices with the National Cable \& Telecommunications Association (NCTA) projecting that the number of connected devices will approximately reach 50 billion devices by the year 2020 \cite{NCTA_IoT,AM_WRV1,AM_WRV2,AM_WRV3,dan_dechene1}. Therefore, a need has risen to manage, process, and store the resulting data being generated more locally. Additionally, new business models, use cases, and applications have emerged that require real-time processing and response. Therefore, the adoption and deployment of MEC infrastructure can help reduce the latency experienced by users. Moreover, it can also help decrease the signaling overhead in the cloud core. This is due to the processing of data being collected and exchanged at the edge nodes rather than at the core.
\subsubsection{Applications}\mbox{}\\
\indent The development and deployment of MEC technologies opened up the door to a variety of services and applications. In what follows, a brief description of how MEC plays a role in some applications is presented.
\begin{enumerate}[a-]
	\item Smart homes: One application that MEC technology is enabling is smart homes. In particular, MEC offers processing and storage capabilities for the large amount of data generated the sensors and various connected appliances within a smart home \cite{edge_for_smart_home,edge_for_smart_home2}. For example, it can help analyze energy consumption patterns based on data collected from smart energy meters \cite{edge_for_smart_home}.  
	\item Smart cities: An extension to the smart homes applications is the smart communities and smart cities application \cite{edge_for_smart_cities}. For example, the video recorded through cameras around the city can be analyzed for routing decisions. Another example is video recorded of a particular monument or location by one user can be cached in a nearby edge node for other users to stream \cite{edge_for_smart_cities}.
	\item Healthcare: A third application for MEC technology is healthcare. For example, remote surgeries can be completed with the help and support of MEC \cite{edge_for_healthcare}. Another example is humanoid robots can offer care-taking services to elderly patients using information collected with MEC technology \cite{edge_for_healthcare}.
	\item Augmented and virtual reality: MEC technologies can play a significant role in supporting augmented and virtual reality (AR/VR) applications \cite{edge_for_ar}. One example is placing a VR control center at a MEC server to improve the tracking accuracy of VR applications \cite{edge_for_ar}.
	\item Retail: Retail is another application that can benefit from MEC technology being deployed. For example, MEC servers can be used to provide Wi-Fi connection to retail store users \cite{edge_for_retail}. MEC servers can also be used to offer smart vending machines and shelves \cite{edge_for_retail}. 
	\item Autonomous and connected vehicles: Another application that MEC technologies can enable is autonomous and connected vehicles within intelligent transportation systems. For example, MEC servers can be used for real-time traffic monitoring and analysis at intersections \cite{edge_for_v2x1,edge_for_v2x2}.
\end{enumerate}
\indent As shown above, MEC technology will play a major role in future networks and applications as it has the potential to support many real-time and latency-sensitive applications. In this work, we focus on the use of MEC in the context of V2X applications and services placement.
\begin{figure*}[!t]
	\centering
	\includegraphics[scale=.6]{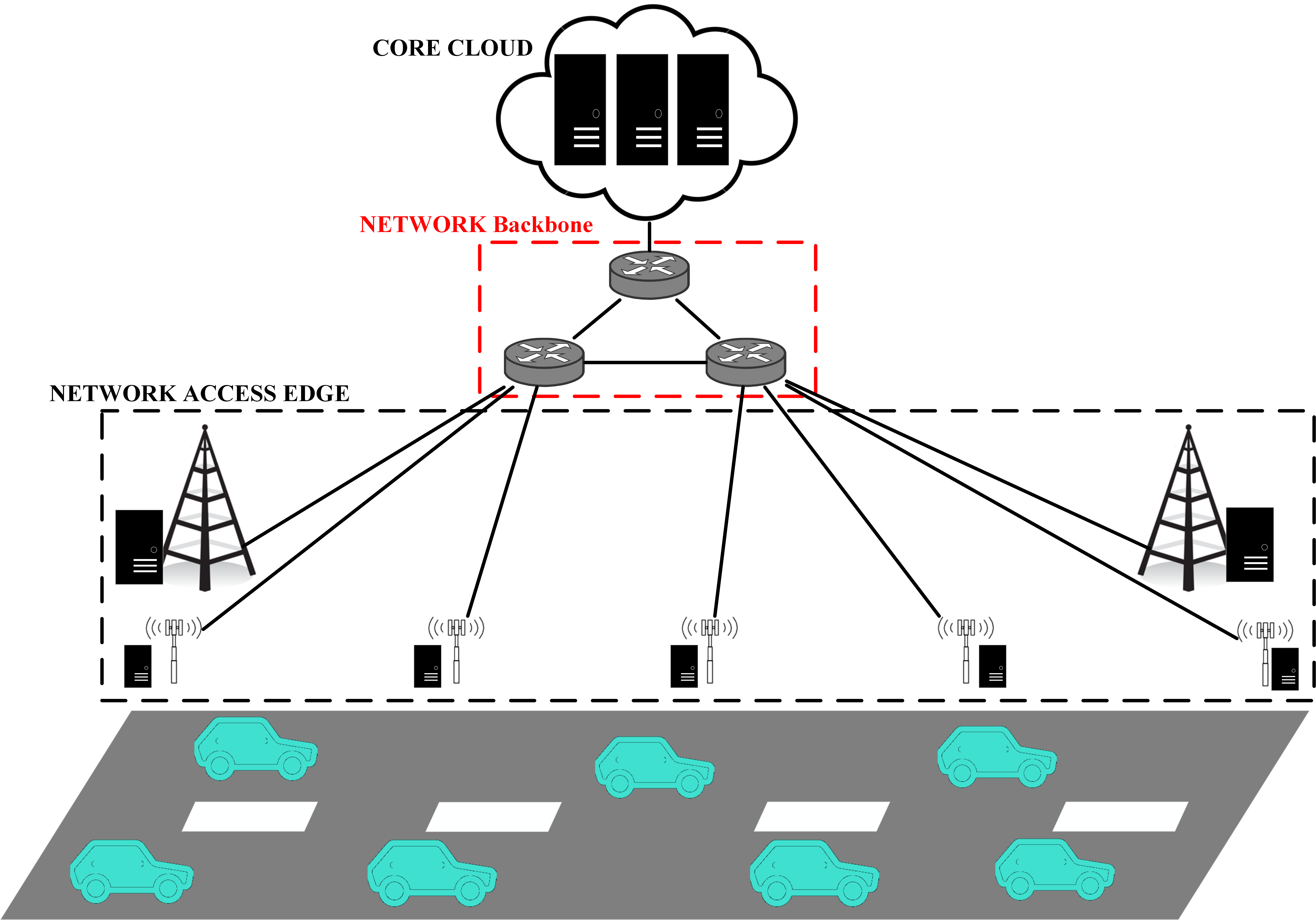}
	\caption{System Model}
	\label{system_model_image}
\end{figure*}
\section{Related Work}\label{related_work}
\indent As mentioned earlier, MEC has been proposed for a variety of applications. Within the context of V2X communication, applications, and services, MEC has been discussed in several previous works. For example, Balid \textit{et al.} proposed the use of MEC technology for real-time traffic surveillance by counting and classifying vehicles \cite{edge_for_v2x1}. The authors developed and implemented a novel low-cost wireless sensor that can be installed on the highway or roadside for traffic monitoring purposes. More specifically, this sensor was able to achieve extremely high vehicle detection accuracy (99.98\%), speed estimation accuracy (97.11\%), and length-based vehicle classification accuracy (97\%).\\
\indent On the other hand, Bissmeyer \textit{et al.} described a security concept to protect V2X communication information and data in a multi-access technology scenario that adopts a MEC-based setup \cite{edge_for_v2x2}. In particular, the authors presented a security framework that ensures message integrity, sender authentication and authorization, and replay detection. This is done through the combined use of digital signatures, public and private key infrastructure, and an Authorization Ticket (AT) certificate. Within this framework, MEC provides local computational power that offers relay event-driven secured V2X messages exchange.\\
\indent Another example of using MEC for V2X applications is the work by Sabella \textit{et al.} in which they proposed a hierarchical MEC architecture for adaptive video streaming \cite{edge_for_v2x3}. In this work, real-time data about the channel conditions is collected by local agents placed at the evolved NodeB (eNB). This data is then shared with a MEC platform that changes the quality of the video stream to match the channel conditions. To test their proposed architecture, the authors implemented a proof-of-concept radio aware video optimization in a virtualized network scenario. The implementation results showed that the proposed architecture achieved a better quality of experience for users through higher downlink and uplink speeds as well as lower latencies.\\
\indent In contrast, Emara \textit{et al.} studied the impact of incorporating MEC on the end-to-end (E2E) latency for cellular-based V2X communication \cite{edge_for_v2x4}. More specifically, the authors focus on how deploying MEC infrastructure can help reduce the E2E latency of the information exchanged between vulnerable road users and the road-vehicles. Through extensive simulations, the authors illustrated that deploying MEC-based infrastructure can help reduce the average E2E latency from around 110 ms to around 25 ms. Moreover, it was shown that the average E2E latency will decrease as the density of vehicle increases when a location-based multi-cast transmission scheme is employed.\\
\indent However, these works have several shortcomings. One such shortcoming is that most of these works only focus on one V2X service or application at a time, be it traffic safety or entertainment. A second shortcoming is that they only study the impact of MEC on metrics such as latency without considering the available computational power at edge nodes. Furthermore, to the best of our knowledge, no other work studied the optimal placement of different V2X services and applications while simultaneously considering the computational resources available, delay requirements, and redundancy requirements of such services/applications. To that end, this paper focuses on formulating the problem of optimal V2X service/application placement in a hybrid core/edge computing environment. .
\section{System Model}\label{sys_model}
\subsection{System Setup}
\indent As shown in Fig. \ref{system_model_image}, this work assumes a highway road environment with multiple lanes in one direction. The vehicles are assumed to be moving with a constant speed along the road with equal distance between them. This can symbolize vehicles along a highway between cities.\\
\indent On the network side, it is assumed that the highway segment in consideration is under LTE-A coverage with multiple evolved NodeB (eNB) base stations. Moreover, there are several RSUs along the road. Each eNB or RSU is equipped with an MEC host having a given set of computing power (CPU, memory, and storage). These eNBs and RSUs form the network access edge. Additionally, the network access edge is assumed to be connected via the network backbone to a core cloud data center that hosts a set of servers with larger computing powers. Note that the LTE-A supports V2X communication through the Uu-based and the PC5-based interfaces \cite{CAM_periodicity,C-V2X3,C-V2X4,D2D-V2X}.
\subsection{V2X Services Description}
\indent This work considers three different types of V2X services. Each of these services represents one use case of some of the previously discussed V2X applications. In what follows, a brief overview of each service and its corresponding requirements is presented.
\subsubsection{Cooperative Awareness Basic Service}\mbox{}\\
\indent The first type of services considered in this work is the Cooperative Awareness Basic Service \cite{CAM1}. This service is characterized by the Cooperative Awareness Message (CAM) that is periodically (typically a 300-byte message every 100ms $\approx$ 3 Kbytes/sec \cite{CAM_periodicity}) shared between the vehicles and the network road nodes. These messages help communicating nodes within a single hop distance to exchange information about their positions, movement, and other sensor information collected \cite{CAM1}. This in turn allows vehicles to make the needed modifications (speed, direction, etc.) based on this information. Accordingly, this service represents a hybrid use case between the autonomous/cooperative driving application and the V2X traffic safety application. To that end, such a service has a stringent latency/delay requirement that is typically between 10-20 ms \cite{messaging_requirement}.    
\subsubsection{Decentralized Environmental Notification Basic Service}
\indent\indent The second type of services this work considers is the decentralized environmental notification basic service \cite{DENM1}. This service is characterized by the Decentralized Environmental Notification Message (DENM) that is typically sent to notify road users of a particular event. Therefore, this message is an event-triggered message (typically around 500 bytes in length and can be sent every 100 ms  $\approx$ 5 Kbytes/sec \cite{CAM_periodicity,DENM_periodicity}) that is sent out to vehicles alerting them to events such as traffic condition warning, road-work warning, and signal violation warning. Therefore, this service represents a hybrid use case between the V2X traffic efficiency and V2X traffic safety applications. Hence, this service has a slightly less stringent latency/delay requirement with latencies up to 100 ms can be tolerated \cite{messaging_requirement}.
\subsubsection{Media Downloading and Streaming}\mbox{}\\
\indent The third type of services explored in this work is media downloading and streaming \cite{media_streaming1}. As the name suggests, this service is one use case of the V2X Infotainment set of applications \cite{media_streaming1}. These applications provide on-demand information or entertainment to vehicle users on either commercial or non-commercial basis \cite{media_streaming1}. This class of services has the least stringent requirement, particularly in terms of delay as it can tolerate up to 1 second of delay for media streaming \cite{media_requirement1,media_requirement2}. This is because such services are not considered to be ``critical'' to driving within an ITS.
\section{Optimized V2X Service Placement (OVSP)}\label{prob_formulation}
\indent This work focuses on solving the problem of optimal placement of V2X services in a hybrid cloud/edge environment. To that end, an analytical optimization model is developed that aims at placing V2X services on computing nodes in such a manner that would minimize the aggregate average delay/latency while satisfying various constraints including delay, computation resource availability, redundancy, and placement constraints. In what follows, the key notations adopted in this work along with a detailed description of the optimized V2X service placement (OVSP) model is given.
\subsection{Key Mathematical Notations:}
\indent The key mathematical notations used in this work are as follows:
\begin{itemize}
	\item $V$: set of vehicles accessing the V2X service instances within communication nodes' coverage range.
	\item $S$: set of V2X services/applications instances to be placed.
	\item $U$: set of unique V2X services/applications.
	\item $S_u$: subset of V2X services/applications of type $u \in U$.
	\item $C$: set of available computing nodes (core node or edge node) to host the V2X services/applications.
	\item $d_{s,v}^{c}$: maximum delay experienced by vehicle $v$ served by V2X service/application instance $s$ if placed at computing node $c$.
	\item $D_{s}^{th}$: maximum tolerable delay/latency of V2X service/application $s$.
	\item $R_{s}^{i}$: computational resource requirement of V2X service $s$, where $i \in \{CPU, memory, storage\}$.
	\item $Cap_{c}^{i}$: maximum available computational resource at computing node $c$.
	\item $Red_{u}^{th};$: minimum redundancy requirement of unique V2X service/application of type $u$. It is assumed that the minimum number of instances that need to be placed is proportional to the number of vehicles within the communication nodes’ coverage range.
\end{itemize}
Accordingly, the decision variable is defined as follows:
\begin{equation}
X_{s}^{c} = \begin{cases}
1, & \text{if V2X service/application $s$ is placed on}\\ 
   & \text{computing node $c$.}\\
0, & \text{otherwise}.
\end{cases}
\end{equation}
\subsection{Problem Formulation:}
\indent Based on the description provided beforehand, the optimal V2X service placement (OVSP) model in a hybrid core/edge computing environment can be formulated as follows:
\begin{figure*}[!tb]
	\centering
	\includegraphics[scale=0.6]{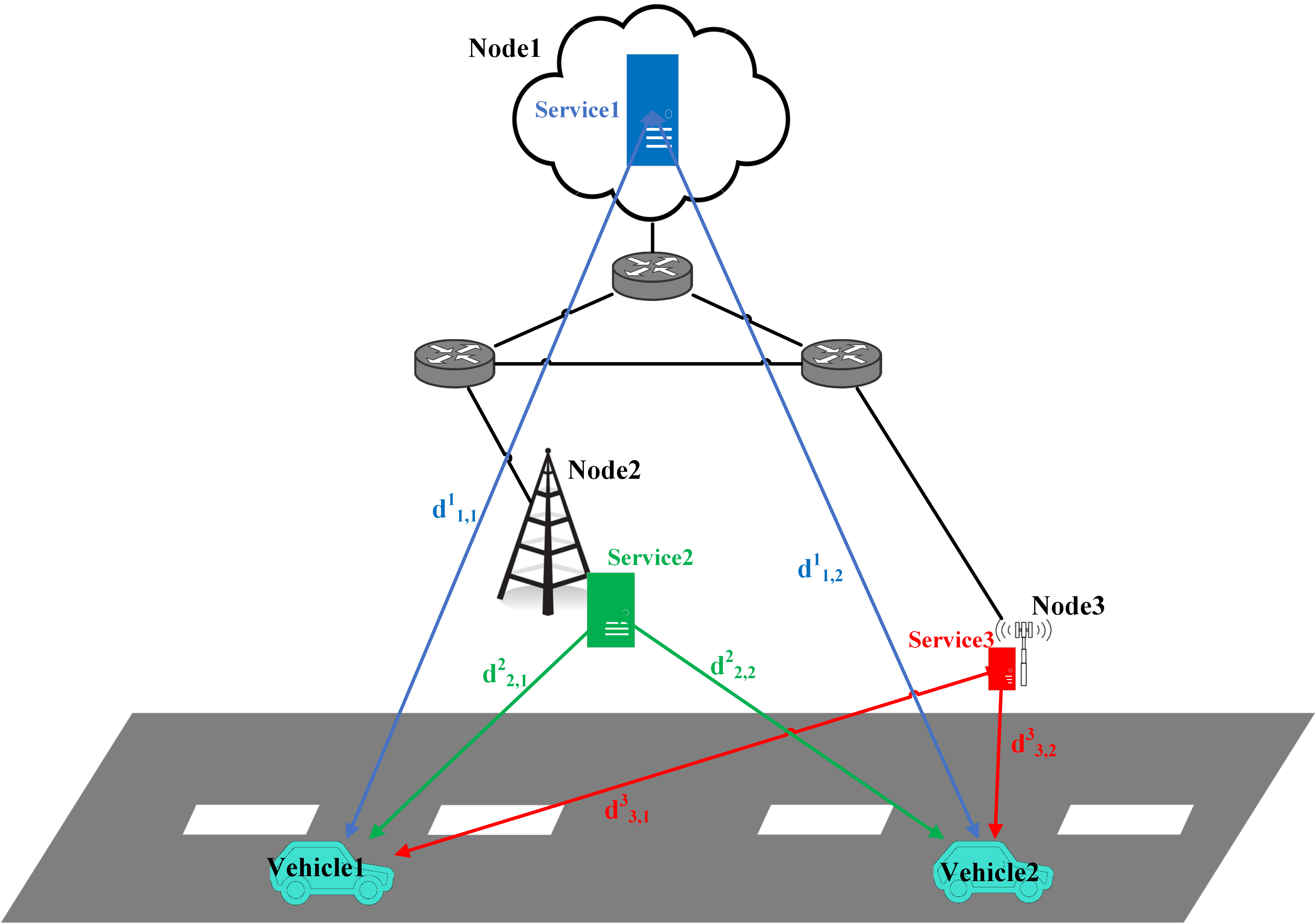}
	\caption{Illustrative Example}
	\label{illustrative_example}
\end{figure*} 
\begin{subequations}\label{problem}
	\begingroup\makeatletter\def\f@size{9}\check@mathfonts
	\begin{equation} \label{objfunc}
	\min \sum\limits_{s \in S}\sum\limits_{c \in C} X_{s}^{c} \left(\frac{1}{|V|}\sum\limits_{v \in V} d_{s,v}^{c}\right)
	\end{equation}
	\indent subject to 
	\begin{equation}\label{delay_constraint}
	\sum\limits_{c \in C} X_{s}^{c} \left( \max\limits_{v \in V} d_{s,v}^{c} \right)\leq D_{s}^{th};\; \forall s \in S
	\end{equation}
	\begin{equation} \label{resources_constraint}
	\sum\limits_{s \in S}  X_{s}^{c} R_{s}^{i} \leq Cap_{c}^{i};\; \forall c \in C, \; \forall i \in \{CPU, memory, storage\}
	\end{equation}
	\begin{equation}\label{redundancy_constraint}
	\sum\limits_{s \in S_u}\sum\limits_{c \in C} X_{s}^{c}  \geq Red_{u}^{th};\; \forall u \in U
	\end{equation}
	\begin{equation}\label{redundant_instances_placement_constraint}
	\sum\limits_{s\in S_u} X_{s}^{c}  \leq 1;\; \forall c \in C;\; \forall u \in U
	\end{equation}
	\begin{equation}\label{placement_constraint}
	\sum\limits_{c \in C} X_{s}^{c} =1 ;\; \forall s \in S
	\end{equation}
	\begin{equation}\label{binary_constraint}
	X_{s}^{c} \in \{0,1\}; \;  \forall s \in S, \; \forall c \in C 
	\end{equation}
	\endgroup
\end{subequations}
\begin{itemize}
	\item Equation (\ref{objfunc}) is the objective function that tries to minimize the aggregate average delay/latency for all the V2X services/applications instances.
	\item Constraint (\ref{delay_constraint}) ensures that the average delay/latency experienced by the vehicles being served by V2X service/application instance $s$ below the service's maximum delay/latency threshold.
	\item Constraint (\ref{resources_constraint}) ensures that the available computational resources at each computing node $c$ are not exceeded.
	\item Constraint (\ref{redundancy_constraint}) ensures the minimum number of instances for unique V2X service/application of type $u$ is placed based on redundancy requirement.
	\item Constraint (\ref{redundant_instances_placement_constraint}) ensures that different instances of unique V2X service/application of type $u$ are placed at different computing nodes.
	\item Constraint (\ref{placement_constraint}) ensures that each instance $s$ is placed at one computing node.
	\item Constraint (\ref{binary_constraint}) specifies that the decision variable $X_{s}^{c}$ is a binary integer decision variable.
\end{itemize}
Based on the above discussion, the problem is considered to be a binary integer linear programming problem. It is worth mentioning that the formulated OVSP model is generic and hence can be applied to any type of V2X service/application placement process. \\
\indent Fig. \ref{illustrative_example} provides an illustrative example of 3 services placed at 3 nodes serving 2 vehicles. The associated delays are also shown. For example, $d_{2,1}^{2}$ represents the delay experienced by vehicle 1 being served by service 2 placed at node 2 (the eNodeB in this case). In this case, Service1 seems to have the highest delay/latency tolerance and hence is placed in the core cloud. In contrast, Service2 and Service3 seem to have more stringent delay/latency requirements and there are placed at edge nodes 2 and 3 respectively. Based on this, the corresponding values for the decision variables would be as follows:
\begin{itemize}
	\item $X_1^1 = X_2^2 = X_3^3 = 1$
	\item $X_2^1 = X_3^1 = X_1^2 = X_3^2 = X_1^3 = X_2^3 = 0$
\end{itemize}
\subsection{Complexity:}
\indent Despite the fact that this problem is a binary integer linear programming problem which can be solved using traditional branch and bound algorithms \cite{bip}, it is well-known that such problems are NP-complete \cite{bip_complexity}. This is supported by the problem's search space. More specifically, the search space for this problem is $2^{|C||S|}$ where $|C|$ is the number of computing nodes and $|S|$ is the number of V2X services to be placed. This is based on the fact that there are $2^{|C||S|}$ possible combinations for the placement of these services on the available computing nodes. For example, for the case of $|C|=10$ and $|S|=10$, the search space is $1.267 \times 10^{30}$. This makes solving such a problem to optimality computationally expensive due to the exponential growth in the search space with the number of services and nodes. Accordingly, a lower-complexity heuristic algorithm needs to be developed to solve this problem.
\section{Greedy V2X Service Placement Algorithm (G-VSPA)}\label{gvspa}
\indent This work proposes a greedy-based heuristic algorithm named ``Greedy V2X Service Placement Algorithm (G-VSPA)'' to solve the V2X service placement problem. Simply put, the algorithm aims at placing each V2X service instance at the closest computing node that has the computing capacity to host it. This is done in an attempt to minimize the aggregate latency to these services. Algorithm \ref{g-vspa} provides the pseudocode of the G-VSPA heuristic algorithm. Note that the proposed G-VSPA algorithm is generic. Accordingly, it can be applied to any type of V2X service placement process. 
\subsection{Description:} 
\indent The algorithm starts by defining a mock variable indicating whether a V2X service instance was placed or not (line 1). It then proceeds to sort the unique services in ascending order of delay tolerance from the least delay-tolerant service to the most delay-tolerant one (line 2). Then the algorithm cycles through the instances of each unique service type. For each instance, the algorithm finds the computing node with the minimum average delay/latency and checks if this node satisfies both the delay and computational resource constraints. If it does, the decision variable is set to 1, the available computing power at node $c$ is updated, and node $c$ is removed from the set of available computing nodes for all other instances of type $u$. This is done to avoid placing two instances of the same type at the same computing nodes. If the constraints are not satisfied, the algorithm moves on to the computing nodes with the second least average delay/latency and checks again. This process repeats itself until all service instances of all unique service types are placed (lines 3-17).
\subsection{Complexity:} 
\indent The time complexity order of this algorithm is $O(|C||S|)$ where $|C|$ is the number of computing nodes and $|S|$ is the number of V2X services to be placed. This is because in each iteration, $|C|$ potential nodes are considered to place each of the V2X service instances available. Therefore, using the same values as in the previous example, the complexity would be in the order of 100 operations. 
\begin{algorithm}[!t]
	\caption{Greedy V2X Service Placement (G-VSPA)}
	\label{g-vspa}
	\begin{algorithmic}[1]
		\renewcommand{\algorithmicrequire}{\textbf{Input:}}
		\renewcommand{\algorithmicensure}{\textbf{Output:}}
		\REQUIRE $U = \{1,2,..,|U|\}$, $S = \{1,2,..,|S|\}$\\ 
		$\;\;\;\;\;C = \{1,2,..,|C|\}$, $V = \{1,2,..,|V|\}$\\
		\ENSURE  $X_{s_u}^{c}$, $Agg\;Delay=\sum\limits_{s \in S}\sum\limits_{c \in C} X_{s}^{c} \left(\frac{1}{|V|}\sum\limits_{v \in V} d_{s,v}^{c}\right)$\\
		\texttt{\\}
		\STATE \textbf{define} $X_{s_u}=\sum\limits_{c \in C} X_{s_u}^{c}$
		\texttt{\\}
		\texttt{\\}
		\STATE\textbf{set} $U_{sort}=Asc\;Sort(U)$
		\texttt{\\}
		\texttt{\\} 
		\FOR{$u \in U_{sort}$}
		\STATE \textbf{define} $C_u=C$
		\FOR{$s_u \in S_u$}
		\WHILE{$X_{s_u}\neq 1$}
		\STATE \textbf{find} $d_{s_u}^c = \min \limits_{c \; \in \; C_u} \{\frac{1}{|V|}\sum\limits_{v \in V} d_{s,v}^{c}\}$
		\texttt{\\}
		\texttt{\\}
		\IF{$Rem\;Cap_c > R_{s_u} \&\;d_{s_u}^c< D_{s_u}^{th}$}{
			\STATE \textbf{set} $X_{s_u}^{c} = 1$
			\STATE \textbf{update} $Rem\;Cap_c\;=Rem\;Cap_c-R_{s_u}$
		\STATE \textbf{update} $C_u \; = C_u \; \backslash \; c$}
		\ELSE{
			\STATE \textbf{update} $d_{s_u}^c\;=\; \infty$
		}
		\ENDIF
		\ENDWHILE
		\ENDFOR
		\ENDFOR
		\RETURN $X_{s_u}^{c}$, $Agg\; Delay$
	\end{algorithmic} 
\end{algorithm}
\section{Performance Evaluation}\label{performance_eval}
\indent To evaluate the performance of the proposed OVSP model and G-VSPA algorithm, a MATLAB-based system level simulator was developed. The performance is evaluated using various metrics such as average delay/latency, delay/latency probability density function, and computational resource utilization. Moreover, the performance of the OVSP model and G-VSPA algorithm is compared to that of the Genetic algorithm (GA), a meta-heuristic that has been previously proposed to solve linear integer programming problems \cite{ga_bip}.
\subsection{Simulation Setup:}
\indent As illustrated in Section \ref{sys_model}, this work adopts an LTE-based V2X system with three different types of services, namely CAM service, DENM service, and media downloading service. The delay tolerance threshold for these services are set at 20, 50, and 150 ms respectively \cite{messaging_requirement,media_requirement1,media_requirement2}. Moreover, it is assumed that the computational requirements of the CAM service are that of a medium sized virtual machine (VM) \cite{service_comp_req,service_comp_req2}. This is mainly because this service is used for local updates of the status of the vehicles. Thus, it has relatively low computational requirement. On the other hand, the DENM service is modeled as large-sized VM as it has to keep track of more data about the traffic conditions based on the information collected through the CAM service \cite{service_comp_req,service_comp_req2}. Therefore, it has higher computational requirements. Last but not least, the media downloading service is modeled as an extra-large VM having the highest computational requirement \cite{service_comp_req,service_comp_req2}. This is because such service needs to keep track and store the various requests for media download for all the vehicles. In terms of the delay/latency experienced by a vehicle, the delay is assumed as follows:
\begin{itemize}
	\item Vehicle-to-RSU Delay: $U(1,10)$ ms \cite{CAM_periodicity}.
	\item Vehicle-to-eNB Delay: $U(20,40)$ ms \cite{CAM_periodicity,access_delay_range}.
	\item Vehicle-to-core Delay: $U(60,130)$ ms \cite{access_delay_range}.
\end{itemize}
\begin{table}[!tp]
	\centering
	\caption{Simulation Parameters}
	\label{Sim_parameter_table}
	\scalebox{0.85}{
		\begin{tabular}{|p{5.5cm}|p{3.8cm}|}
			\hline
			\textbf{Parameter} & \textbf{Value} \\ \hline
			Number of lanes & 2 \\ \hline
			Core node computing resources\newline [CPU, Memory, Storage]& [32 cores, 64 GB, 240 GB]\\ \hline
			eNB node computing resources\newline [CPU, Memory, Storage]& [8 cores, 16 GB, 240 GB]\\ \hline
			RSU node computing resources\newline [CPU, Memory, Storage]& [8 cores, 16 GB, 240 GB]\\ \hline
		    \vspace{0.1cm}Service Delay Threshold\newline [CAM, DENM, Media downloading]&\vspace{0.1cm} [20 ms, 50 ms, 150 ms]\\ \hline
			\vspace{0.1cm}CAM service computing requirement \newline [CPU, Memory, Storage]&\vspace{0.1cm} [2 cores, 3.5 GB, 4 GB]\\ \hline
		    \vspace{0.1cm}DENM service computing requirement\newline  [CPU, Memory, Storage]&\vspace{0.1cm}[4 cores, 7 GB, 4 GB]\\ \hline
			\vspace{0.1cm}Media download service computing \newline requirement [CPU, Memory, Storage]&\vspace{0.1cm}[8 cores, 14 GB, 40 GB]\\ \hline
			\multicolumn{2}{|c|}{\textbf{Scenario 1: Small Scale Scenario}}\\ \hline
			Lane length (km) & 2 \\ \hline
			Number of vehicles & [20, 40, 60, 80, 100]\\ \hline
			Number of core nodes & 2\\ \hline
			Number of eNB nodes & 3\\ \hline
			Number of RSU nodes & 5\\ \hline
			Number of CAM instances & [1-5]\\ \hline
			Number of DENM instances & [1-5]\\ \hline
 			Number of Media instances & [1-5]\\ \hline			
			\multicolumn{2}{|c|}{\textbf{Scenario 2: Large Scale Scenario}}\\ \hline
			Lane length (km) & 8 \\ \hline
			Number of vehicles & [140, 180, 220, 260, 300]\\ \hline
			Number of core nodes & 7\\ \hline
			Number of eNB nodes & 8\\ \hline
			Number of RSU nodes & 15\\ \hline
		    Number of CAM instances & [7-15]\\ \hline
			Number of DENM instances & [7-15]\\ \hline
			Number of Media instances &[7-15]\\ \hline
	\end{tabular}}
\end{table} 
\indent This is done for two different scenarios, namely a small scale scenario and a large scale scenario. This is done to verify any observed trends. Table \ref{Sim_parameter_table} summarizes the simulation parameters describing the physical environment for both scenarios in terms of the number of vehicles, number of nodes, and the computational resources available at these nodes \cite{cloud_server_specs,edge_server_specs1,edge_server_specs2}. The results are averaged over 100 runs for each value.
\subsection{Results and Discussion:}
\subsubsection{Small Scale Scenario}
\begin{figure}[!t]
	\centering
	\includegraphics[scale=0.075]{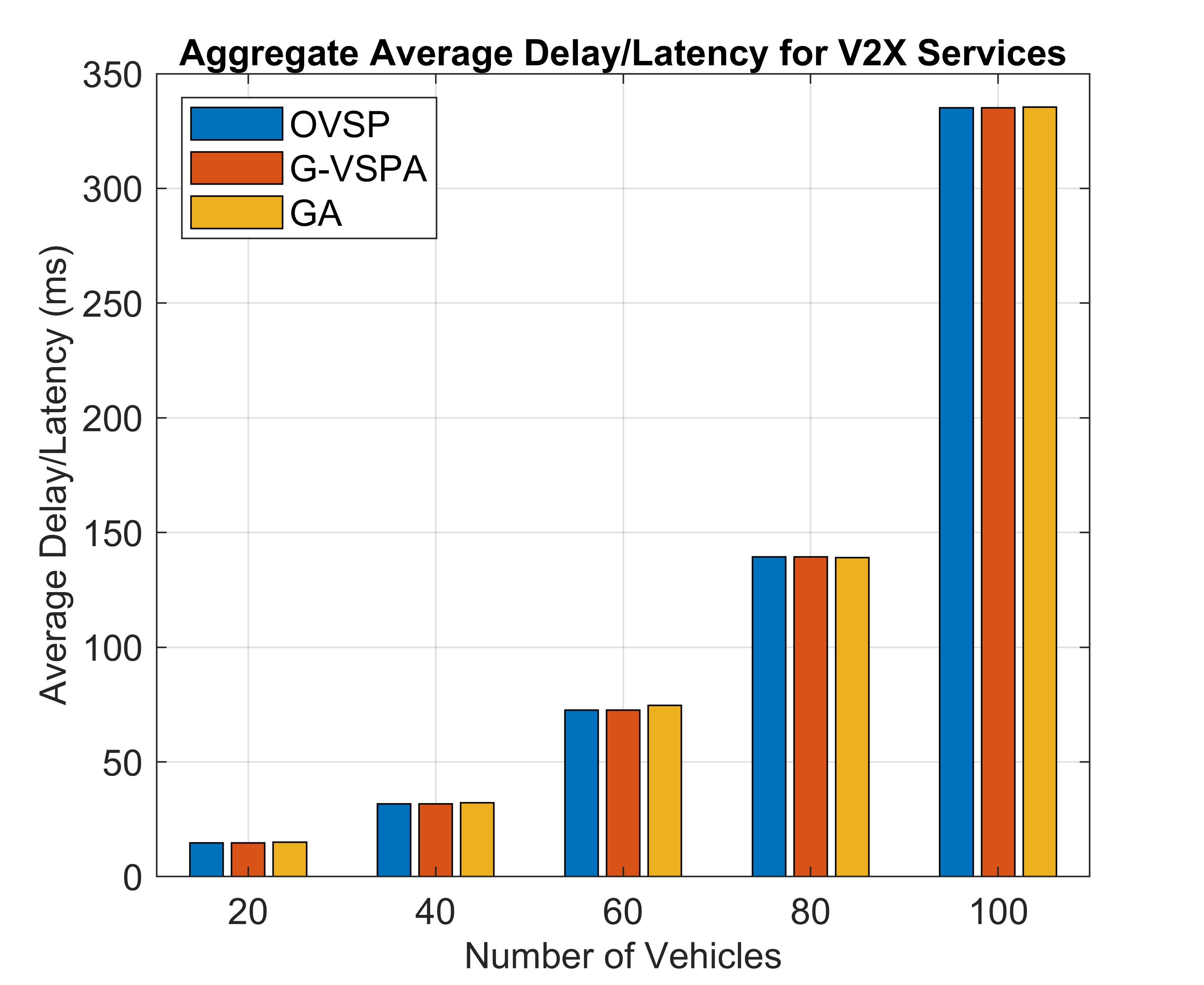}
	\caption{Scenario 1: Aggregate Average Delay/Latency}
	\label{Agg_avg_delay}
\end{figure}

\begin{figure}[!tb]
	\centering
	\includegraphics[scale=0.075,trim=0cm 1cm 0cm 0cm]{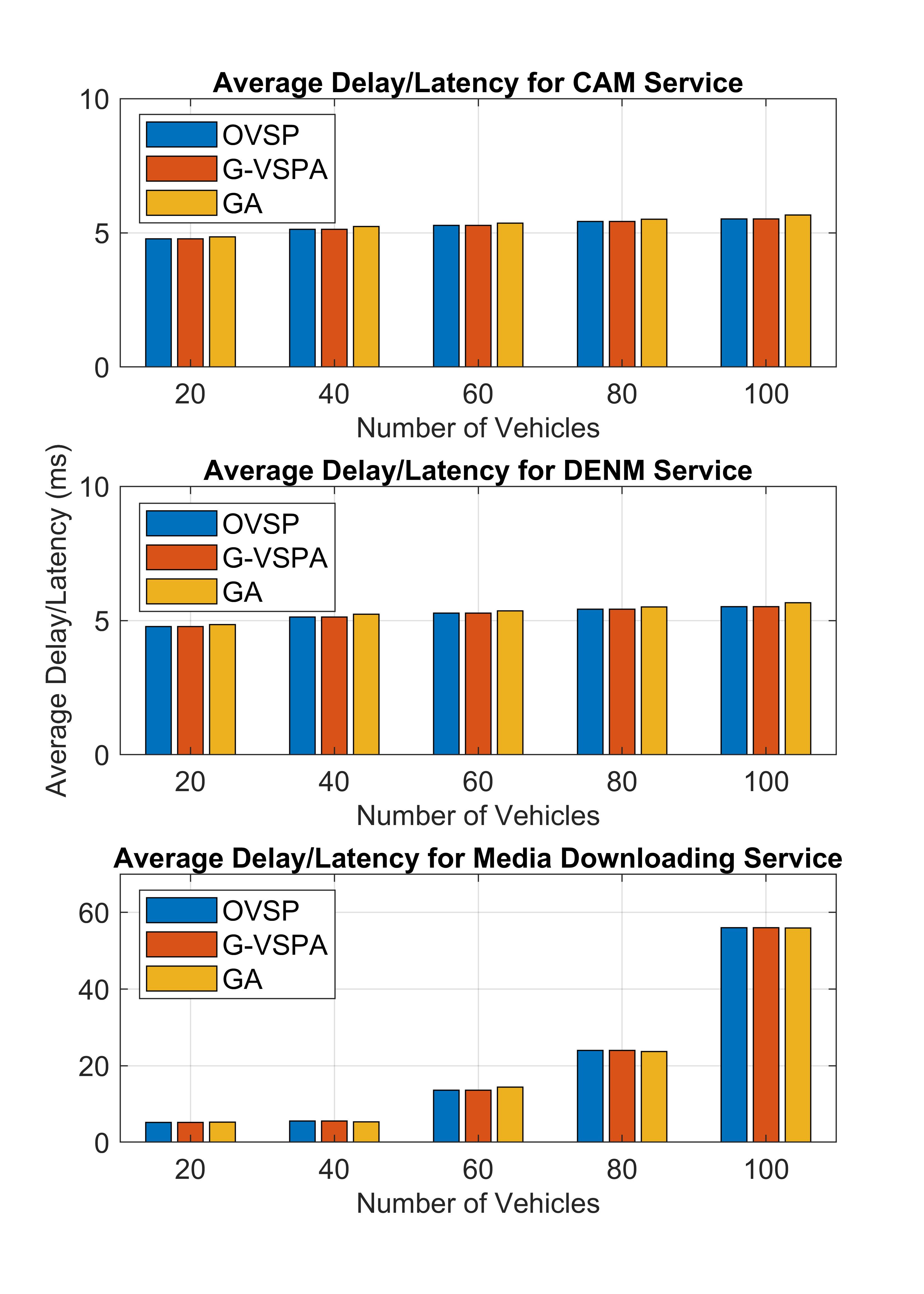}
	\caption{Scenario 1: Average Delay/Latency For Different V2X Services}
	\label{avg_delay_service}
\end{figure}

\begin{figure}[!tb]
	\centering
	\includegraphics[scale=0.075]{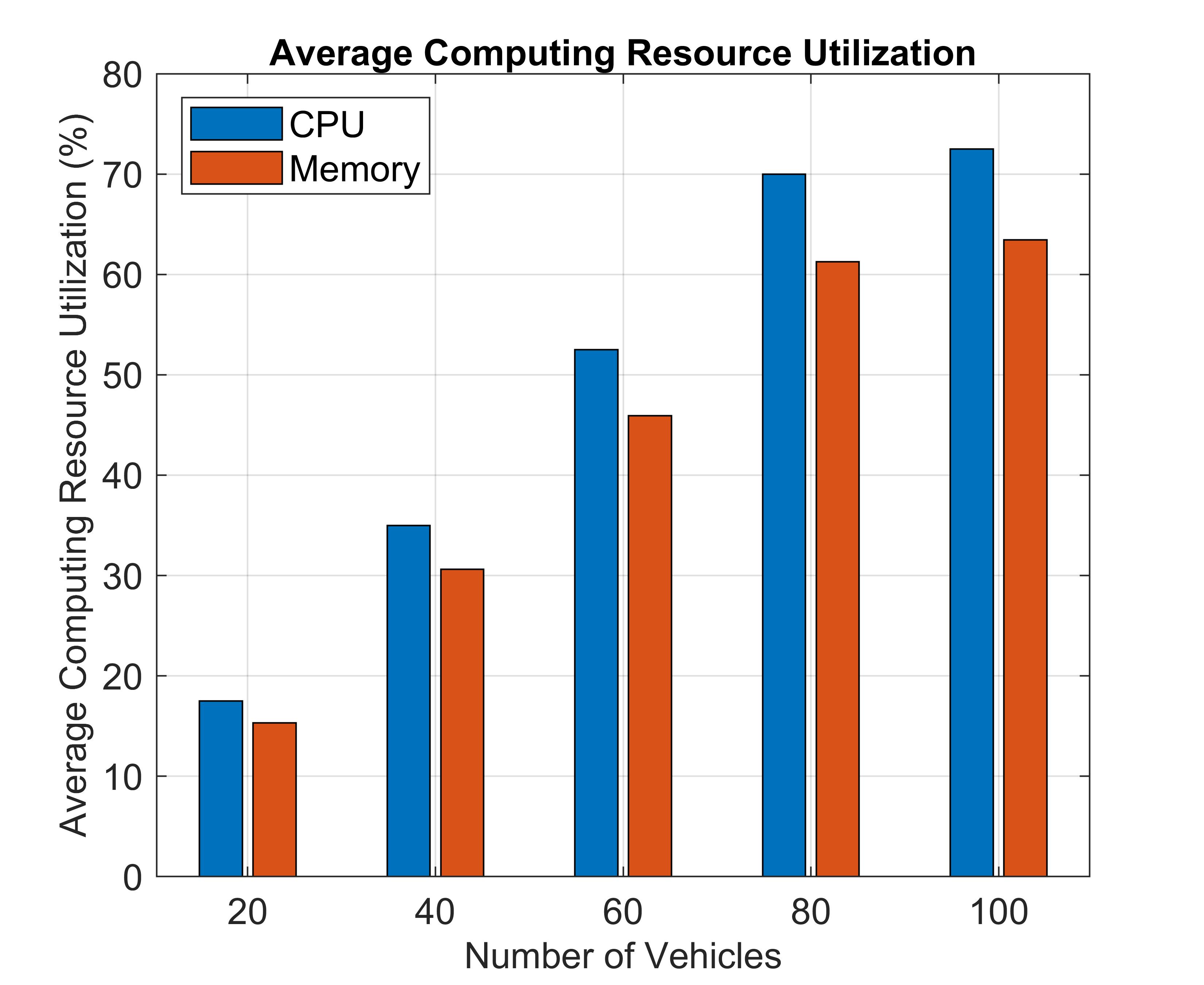}
	\caption{Scenario 1: Average Computing Resource Utilization}
	\label{computing_util}
\end{figure}

\begin{figure}[!tb]
	\centering
	\includegraphics[scale=0.075,trim=0cm 1.5cm 0cm 0cm]{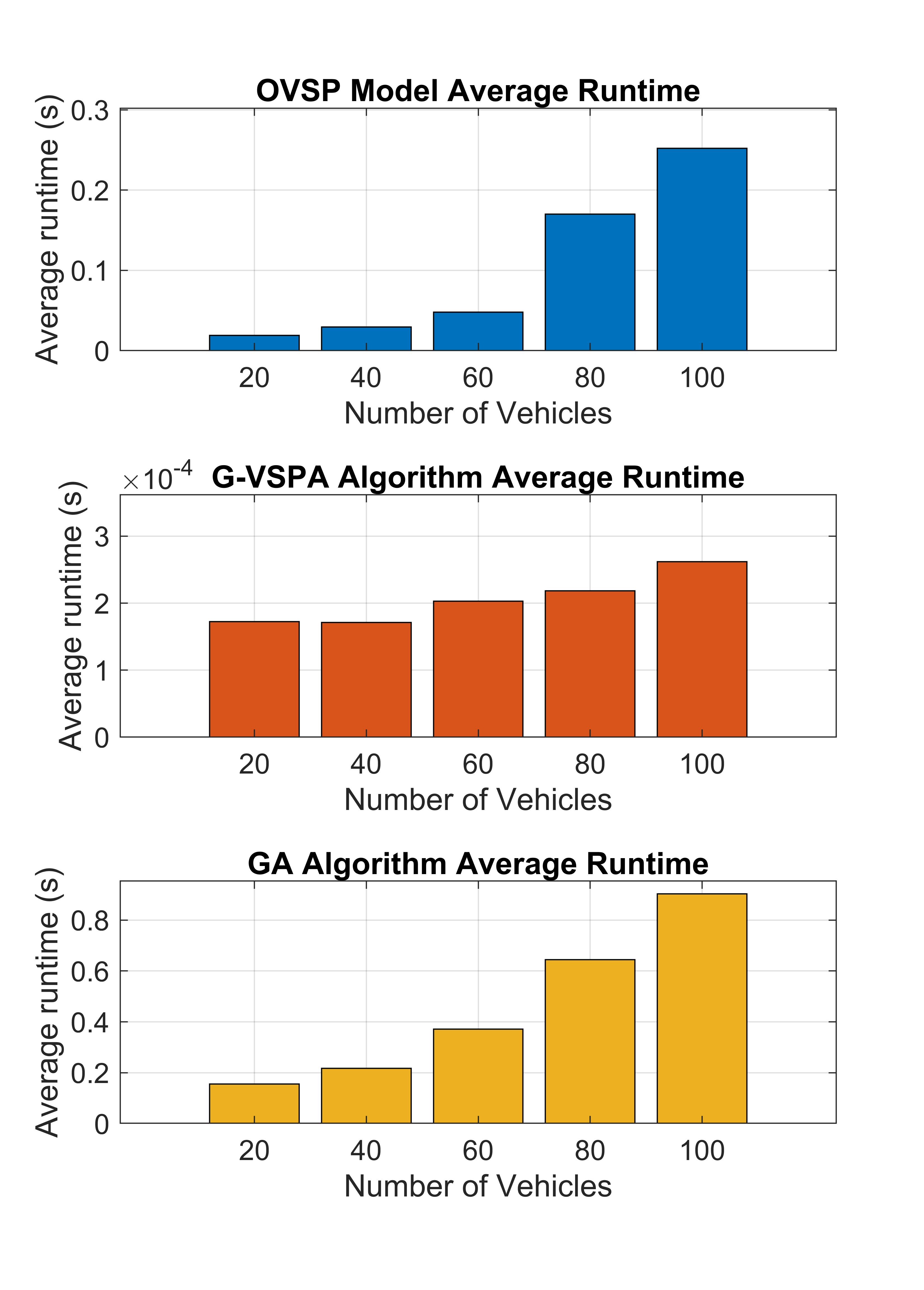}
	\caption{Scenario 1: Average Runtime For OVSP Model (top), G-VSPA Algorithm (middle), and GA Algorithm (bottom)}
	\label{avg_runtime_fig}
\end{figure}

\mbox{}\\
\indent Fig. \ref{Agg_avg_delay} illustrates the aggregate average delay/latency. As expected, the aggregate average delay/latency increases as the number of vehicles increases. This is because more V2X service instances need to be placed to handle the growing number of vehicles. Additionally, more instances are placed at the core due to the limited computing power available at the edge nodes, resulting in higher average delays/latencies. Another observation is that the proposed G-VSPA heuristic algorithm achieves close to optimal performance when compared to both the OVSP model and GA algorithm. This is due to the fact that the heuristic also tries to place the services at the edge and moves to the core as the number of instances to be placed increases. In contrast, the GA algorithm is prone to giving inefficient solutions in some cases due to an unsuitable initialization of the population. In turn, this can lead to having a placement solution with a higher average latency. Moreover, this figure shows that both the OVSP model and G-VSPA algorithm provide a suitable placement for the V2X services. This is based on the fact that the aggregate average delay/latency is below the threshold for 80 vehicles, meaning that the delays/latencies experienced by each service instance is well below its threshold. These results further highlight the effectiveness of the proposed G-VSPA algorithm as it performs close to optimal while having a significantly lower complexity. 
\indent Building on the aforementioned results, Fig. \ref{avg_delay_service} shows the average delay/latency for each V2X service type. It can be observed that the average delay/latency for CAM and DENM services remain stable while that of the media downloading service increases as the number of vehicles increases. This is due to the fact that the CAM and DENM services have more stringent delay requirements than media downloading. This results in such services being placed at the edge. In contrast, some of the media downloading service instances are placed at the core cloud as it has higher computing power and can still satisfy the service's delay requirement. Note that these observations hold true for the OVSP model, G-VSPA algorithm, and GA algorithm.\\
\indent Fig. \ref{computing_util} shows the average CPU and memory utilization of a computing node as the number of vehicles increases. As can be seen, the average computing resource utilization increases as the number of vehicles increases. This is due to the increased number of V2X service instances that need to be placed to handle the connected vehicles. Additionally, due to the limited computational resources available at the edge, such nodes will be saturated quickly, resulting in higher utilization values. \\
\indent Fig. \ref{avg_runtime_fig} shows the average runtime for the OVSP model, the G-VSPA algorithm, and the GA algorithm respectively. Several observations can be made. The first is that the average runtime is much lower for the G-VSPA algorithm when compared to the OVSP model and the GA algorithm. This is evident by the fact that the average runtime for the G-VSPA algorithm was close to 0.2ms while that of the OVSP model was around 70 ms and that of the GA algorithm was close to 450 ms. This is expected given the lower order of complexity of the G-VSPA algorithm when compared to that of the OVSP model and the GA algorithm (whose complexity is dependent not only on the problem size, but also the population size \cite{ga_complexity}). Moreover, it can be observed the runtime increases as the number of vehicles increases for the OVSP model, but remains more stable for the G-VSPA model. This further highlights the benefit of the G-VSPA algorithm as it achieves close to optimal performance with a significantly lower runtime.   

\subsubsection{Large Scale Scenario}
\begin{figure}[!t]
	\centering
	\includegraphics[scale=0.075]{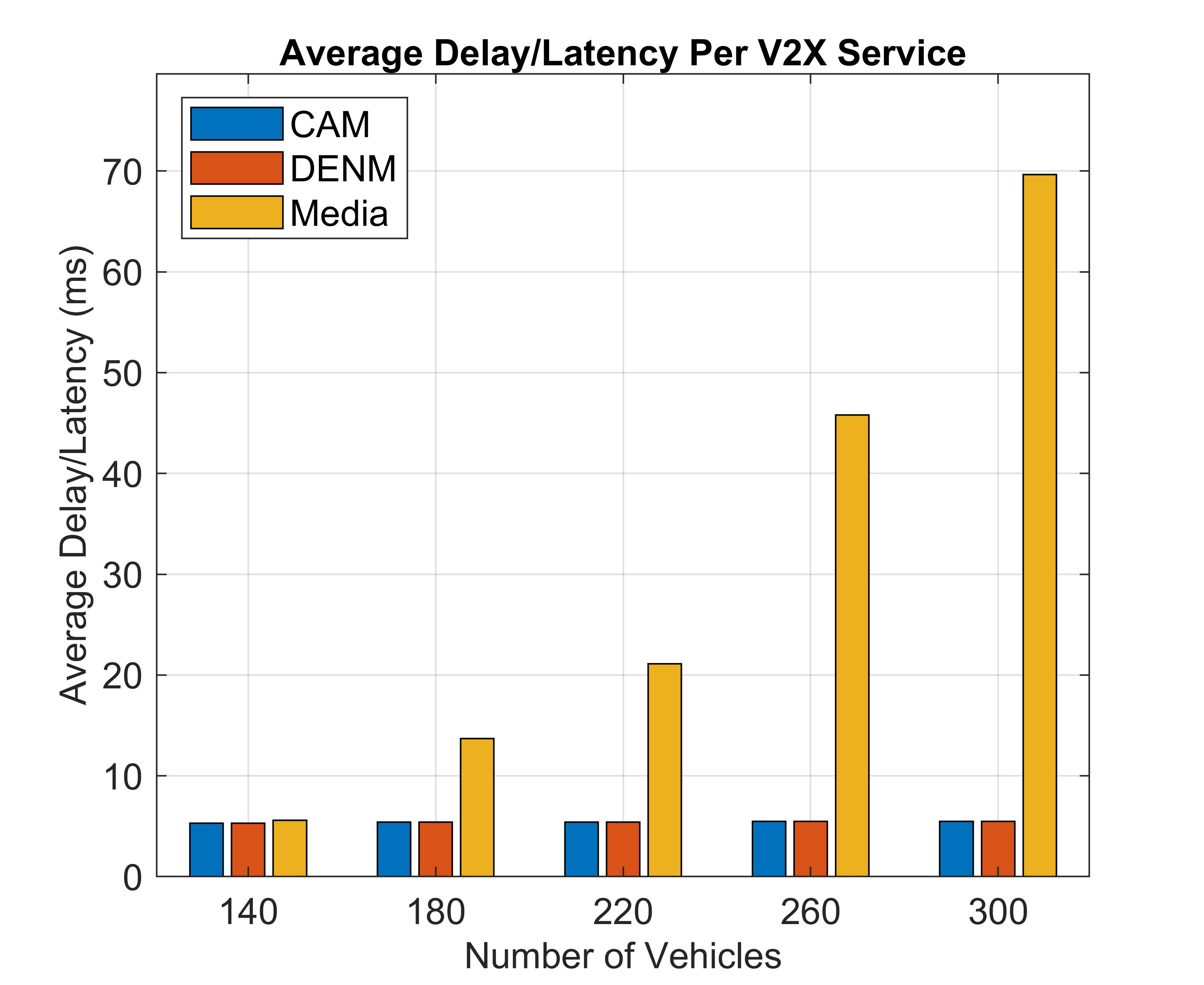}
	\caption{Scenario 2: Average Delay/Latency For Different V2X Services}
	\label{avg_delay_service_s2}
\end{figure}

\mbox{}\\
\indent To verify the observed trends in the small scale scenario, a larger environment is considered with a longer lane length, more vehicles, and a larger number of available computing nodes. It is worth mentioning that only the G-VSPA was implemented in Scenario 2. This is due to two main reasons. The first is the time complexity associated with solving the OVSP model for such a large scenario. The second is due to the potentially inaccurate results that the GA algorithm can give due to unsuitable population initialization in some cases, an issue that can be exacerbated in a large-scale scenario. However, as shown previously, the G-VSPA results provide a good estimate of the performance of the OVSP model.\\
\indent Fig. \ref{avg_delay_service_s2} shows the average delay/latency for the different V2X services. Again, it can be observed that the average delay/latency for CAM and DENM services remain stable while that of the media downloading services will increase as the number of vehicles increases. This proves that due to the stringent delay/latency requirements of services such as CAM and DENM, they will always be placed at the edge. In contrast, media downloading services can be placed at both edge and core nodes due to their higher delay/latency tolerance. 

\begin{figure}[!t]
	\centering
	\includegraphics[scale=0.075,trim=0cm 1cm 0cm 0cm]{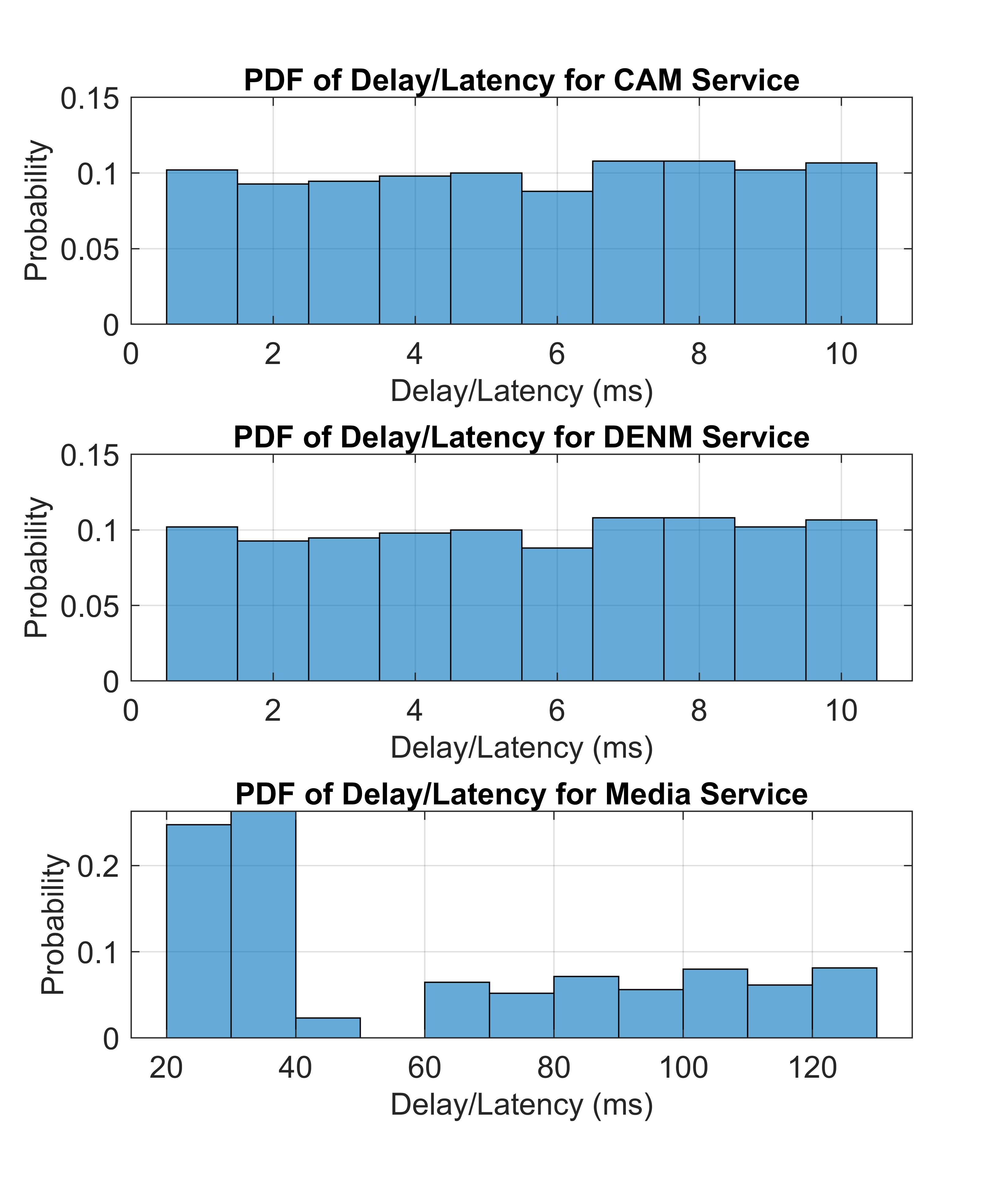}
	\caption{Scenario 2: PDF of Delay/Latency For Different V2X Services}
	\label{delay_pdf}
\end{figure}

\indent To further verify the previous results, the probability density function of the delays/latencies experienced by vehicles for both the CAM and media downloading services is shown in Fig. \ref{delay_pdf}. This figure shows that the delay/latency experienced by vehicles for the CAM services is always below the 20 ms threshold. This further cements the notion that such services will always be placed at edge nodes. On the other hand, the delay/latency experienced by vehicles for media downloading service instances range between 20 and 130 ms, which means that instances of this service were placed at both edge and core nodes. However, it can still be observed that the delay is below the threshold of 150 ms for such services.
\section{Conclusion}\label{conc}
\indent Vehicle-to-everything (V2X) communication and services are playing a major role in future intelligent transportation systems (ITSs). However, many of these services have stringent performance requirements, particularly in terms of the delay/latency. To address this issue, multi-access/mobile edge computing (MEC) has been proposed as a potential solution by placing such services at edge nodes to bring them closer to vehicles. Yet, this results in a new set of challenges. One such challenge is  where to place these V2X services, especially given the limit computation resources available at edge nodes. To that end, this work formulated the problem of optimal V2X service placement (OVSP) in a hybrid core/edge environment as a binary integer linear programming problem. Additionally, a low-complexity greedy-based heuristic algorithm, namely ``Greedy V2X Service Placement Algorithm'' (G-VSPA), was developed to solve this problem. Using extensive simulation, results showed that the OVSP model successfully maintained and satisfied the QoS requirements of all the different V2X services by placing delay-stringent services at the edge and delay-tolerant services at the core nodes. These trends were verified for two simulation scenarios. Additionally, it was shown that the proposed G-VSPA algorithm achieved close to optimal performance while having lower computational complexity.\\ 
\indent To further extend this work, it is worth exploring the impact of cost-efficient V2X placement on the performance of the system. This is because placing services at the core tends to be more cost-efficient. Hence, investigating the trade-off between delay performance and cost needs should be investigated. Moreover, compound services with higher network throughput requirements should be considered by adapting the corresponding optimization model  and heuristic algorithm. This is crucial given that many of the compound services (made up of a group of basic services) have a high network throughput requirement. Therefore, studying the trade-off between delay and throughput is a research avenue worth exploring. 
 
\small
\bibliographystyle{IEEEtran}
\bibliography{References}
\begin{IEEEbiography}[{\includegraphics[width=1in,height=1.25in,clip,keepaspectratio]{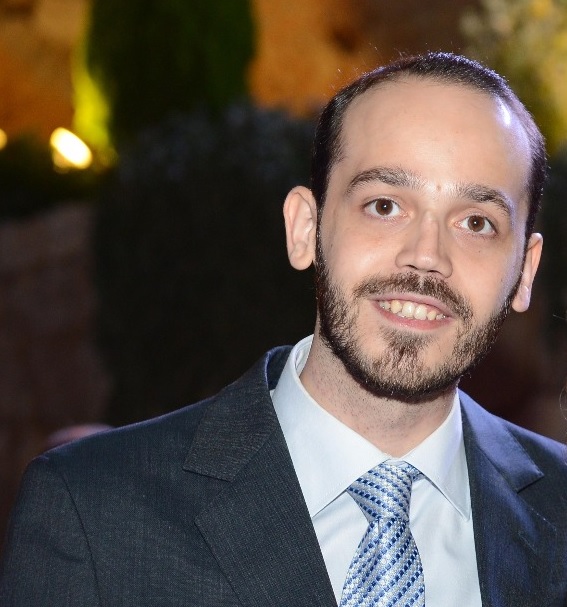}}]{Abdallah Moubayed}
	received his Ph.D. in Electrical \& Computer Engineering from the University of Western Ontario in August 2018,  his M.Sc. degree in Electrical Engineering from King Abdullah University of Science and Technology, Thuwal, Saudi Arabia in 2014, and his B.E. degree in Electrical Engineering from the Lebanese American University, Beirut,
	Lebanon in 2012. Currently, he is a Postdoctoral Associate in the Optimized Computing and Communications (OC2) lab at University of Western Ontario.  His research interests include wireless communication, resource allocation, wireless network virtualization, performance \& optimization modeling, machine learning \& data analytics, computer network security, cloud computing, and e-learning. 
\end{IEEEbiography}	

\begin{IEEEbiography}[{\includegraphics[width=1in,height=1.25in,clip,keepaspectratio]{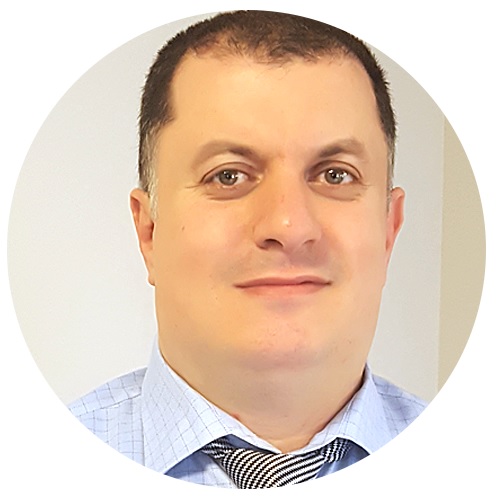}}]{Abdallah Shami}
	is a Professor at the ECE department at Western University, Ontario, Canada. Dr. Shami is the Director of the Optimized Computing and Communications Laboratory at Western. He is currently an Associate Editor for IEEE Transactions on Mobile Computing, IEEE Network, and IEEE Communications Tutorials and Survey. Dr. Shami has chaired key symposia for IEEE GLOBECOM, IEEE ICC, IEEE ICNC, and ICCIT. He was the elected Chair of the IEEE Communications Society Technical Committee on Communications Software (2016-2017) and IEEE London Ontario Section Chair (2016-2018). 
\end{IEEEbiography}	

\begin{IEEEbiography}[{\includegraphics[width=1in,height=1.25in,clip,keepaspectratio]{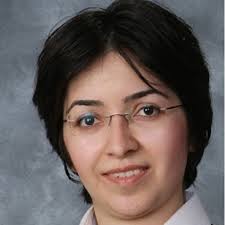}}]{Parisa Heidari}
	is working as a Software Engineer at Edge Gravity by Ericsson. She received her Masters and PhD in computer engineering from Ecole Polytechnique de Montreal, Canada in 2007 and 2012, respectively. She worked as research associate at Concordia University in collaboration with Ericsson from 2013-2014. In 2015, she joined Ericsson Research in Montreal as a postdoctoral fellow. She is working at Edge Gravity by Ericsson since 2017. She holds to her credit several publications and patents. Her research interests include Edge Computing, Cloud next generation, container technologies and server-less approach, smart resource dimensioning, optimal placement and different aspects of QoS assurance in cloud systems.
\end{IEEEbiography}	

\begin{IEEEbiography}[{\includegraphics[width=1in,height=1.25in,clip,keepaspectratio]{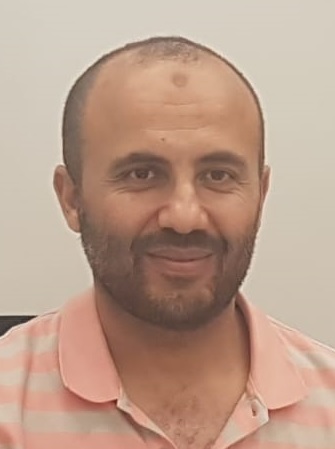}}]{Adel Larabi}
is a Senior Solution Architect at EDGE GRAVITY  by Ericsson with over 25 years of leadership experience in designing innovative business solutions for Telco. Helping bridging academia research projects with commercial grade enterprise solutions. Core qualifications in CDN, Edge Computing, Big data, IMS, Media, and OSS with interest on AI applied to these domain.
\end{IEEEbiography}	

\begin{IEEEbiography}[{\includegraphics[width=1in,height=1.25in,clip,keepaspectratio]{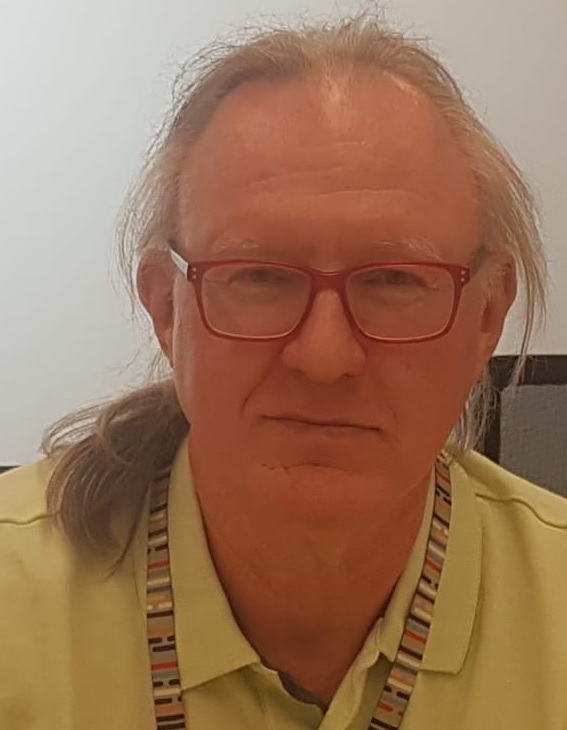}}]{Richard Brunner}
joined Ericsson in 1988 and with broad career experience in System Management, Standardization and Strategic Product management. Richard possesses international management experience, with deep knowledge of the wireless telecommunication market. Richard is actively engaged in setting  Ericsson's  5G TV and EdgeGravity  research and partnership activities both internally and externally towards industry and universities.
\end{IEEEbiography}	
\end{document}